\documentclass[A4paper, 11pt]{article}
\usepackage{}
\usepackage{amsfonts}
\usepackage{dsfont}
\usepackage{mathrsfs}
\usepackage{amssymb}
\usepackage{bbm}
\usepackage{epsfig}
\usepackage{amsmath}
\usepackage{appendix}
\usepackage{float}
\usepackage{color}
\usepackage[english]{babel}

\def\Reals{\mathop{\hbox{\mit I\kern-.2em R}}\nolimits}

\def\Complexes{{\hbox{\mit C\kern-.46em
            \vrule depth 0ex height 1.4ex width .05em\kern.41em}}}

\newtheorem{thm}{Theorem}
\newtheorem{defn}{Definition}

\newtheorem{lem}{Lemma}
\newtheorem{remark}{Remark}
\newtheorem{prop}{Proposition}

\title{\bf How Agreement and Disagreement Evolve over Random Dynamic
  Networks\footnote{This work has been supported in part by the Knut
    and Alice Wallenberg Foundation, the Swedish Research Council and  KTH SRA TNG.}}
\date{}
\author{Guodong Shi, Mikael Johansson, and Karl H. Johansson\thanks{The authors are with ACCESS Linnaeus Centre, School of Electrical Engineering,
Royal Institute of Technology, Stockholm 10044, Sweden.
       Email: {\tt\small $\{$guodongs, mikaelj,  kallej$\}$@kth.se}}}
\begin{document}
\maketitle
\begin{abstract}
The dynamics of an agreement protocol interacting with a disagreement process over a common random network is considered. The model can represent the spreading of true and false information over a communication network, the propagation of faults in a large-scale control system, or the development of trust and mistrust in a society. At each time instance and with a given probability, a pair of network nodes are selected to
interact. At random each  of the nodes then updates its state towards
the state of the other node (attraction), away from the other node
(repulsion), or sticks to its current state (neglect). Agreement convergence
and disagreement divergence results are obtained for various strengths of the
updates for both symmetric and asymmetric update rules. Impossibility
theorems show that a specific level of attraction is required for
almost sure asymptotic agreement and a specific level of repulsion is
required for almost sure asymptotic disagreement. A series of
sufficient and/or necessary conditions are then established for
agreement convergence or disagreement divergence. In particular, under
symmetric updates, a
critical convergence measure in the attraction and repulsion update
strength is found, in the sense that the asymptotic property of the network state evolution transits from
agreement convergence to disagreement divergence when this measure goes from negative to
positive. The result can be interpreted as a tight bound on
how much bad action needs to be injected in a dynamic network in order
to consistently steer its overall behavior away from consensus.
\end{abstract}


\section{Introduction}

\subsection{Motivation}
A growing number of applications are composed of a networked information structure executed over an underlying communication network. Examples include social networks run over the Internet, control networks for the power grid, and information networks serving transportation systems. These networks are seldom centrally regulated, but have a strong component of distributed information processing and decision-making. While these systems are able to provide appropriate service to their users most of the time, open software and communication technologies, together with the large geographical scale of the systems, make them more exposed to faulty components, software bugs, communication failures and even purposeful injection of false data.

An interesting problem is to try to understand the amount of deficiencies that can be tolerated in the combined network before the global system performance is compromised. In this paper we tackle this challenging problem for a model inspired by agreement protocols, whose execution have been studied intensively over the last decade in a variety of settings, including load balancing in parallel computing~\cite{cs2,cs3}, coordination of
autonomous agents~\cite{saber04,mar}, distributed estimation and signal
processing~\cite{moura1,moura2}, and opinion dynamics in social
networks~\cite{social1,social2,social3}. In this model, a pair of nodes is selected randomly at each time. The nodes update their scalar states by forming a weighted average of their own state with the state of the other node. Such a dynamic network protocol is sometimes called gossiping and its convergence is fairly well understood. To account for system defects and node misbehavior, we extend the basic gossiping model in the following way. Rather than always executing the regular update, which we call an attraction step, nodes do so with a certain probability every time they are drawn. If a node does not perform the attraction step, it randomly either updates its own state away from the other node's state (we call this a repulsion step) or simply chooses not to perform a state update but to stick to its current value (which we call neglect).  We believe that our model is one of the first to consider faulty and misbehaving nodes in gossiping algorithms. Based on our model, it is possible to analyze almost sure (a.s.) convergence to agreement and derive explicit criteria for the amount of node deficiencies that can be tolerated while still guaranteeing that all node states converge to a common value. By interpreting the repulsion step as a simple model for node misbehavior or faults, these criteria also characterize the strength or persistency of node misbehavior that is required to force the nodes to an overall disagreement.
\subsection{Related Work}
The structure of complex networks, and the dynamics of the internal states of the nodes in these networks, are two fundamental issues in the study of network science \cite{rg2,rg3}.

Probabilistic models for networks such as random graphs, provide an important and convenient means for modeling  large-scale systems, and have found numerous applications in various fields of science.   The classical Erd\"{o}s--R\'{e}nyi model,  in which each edge exists randomly and independently of others with a given probability, was studied in \cite{er}. The degree distribution of the Erd\"{o}s--R\'{e}nyi graph is asymptotically Poisson. Generalized models were proposed in \cite{watts} and \cite{bara}, for which the degree distribution satisfies  certain power law that better matches the properties of real-life networks such as the Internet. A detailed introduction to the structure of random networks can be found in \cite{rg1,rg2}.

When a networked information  is executed on top of an underlying network, nodes are endowed with internal states that evolve as nodes interact. The dynamics of the node states depend on the particular problem under investigation. For instance, the boids model was introduced in  \cite{boid} to model swarm behavior and animal groups, followed by Vicsek's model in \cite{vic95}.    Models of opinion dynamics in social networks were considered in \cite{social1,social2,daron} and the dynamics of communication protocols in  \cite{shah}. Distributed averaging or consensus algorithms have relative simple dynamics for the network state evolution and serve as a basic model for the complex interaction between node state dynamics and the dynamics of the underlying communication graph.

Convergence to agreement for averaging algorithms have been extensively studied in the literature. Early results were developed in a general setting for studying the ergodicity of nonhomogeneous Markov chains~\cite{haj,wolf}. Deterministic models  have been investigated  in
finding proper connectivity conditions that ensure consensus
convergence~\cite{tsi, jad03,  tsi2,
  caoming1, caoming3, mor,  ren, julien1,julien2}. Averaging algorithms over random graphs have also been considered~\cite{hatano, wu,
  jad2,fagnani1, fagnani2, bamieh, boyd1}.

In this paper,  we use the asynchronous time model introduced in \cite{boyd} to describe the randomized  node  interactions. Each node meets other nodes at independent time instances  defined  by a  rate-one Poisson process, and then a pair of nodes is selected to meet at random determined by the  underlying communication graph.  Gossiping, in which each
node is restricted to exchange data and decisions with at most one
neighboring node at each time instance, has proven to be a robust and
efficient way to implement distributed computations and signal
processing~\cite{ moura2, gossip2, gossip1, boyd,
  f-z,bdo,murray,daron, shah}. A central problem here is to analyze
if a given gossip algorithm converges to consensus, and to determine
the rate of convergence of the consensus process. Karp et
al.~\cite{gossip2} derived a general lower bound for synchronous
gossip. Kempe et al.~\cite{gossip1} proposed a randomized gossip
algorithm on complete graphs and determined the order of its
convergence rate. Boyd et al.~\cite{boyd} established both lower
and upper bounds for the convergence time of synchronous and
asynchronous randomized gossip algorithm, and developed algorithms
for optimizing parameters to obtain fast consensus. Fagnani and
Zampieri discussed asymmetric gossiping in \cite{f-z}. Liu et
al.~\cite{bdo} presented a comprehensive analysis for the asymptotic
convergence rates of deterministic averaging, and recently
distributed gossip averaging
 subject to quantization constraints was studied in~\cite{murray}. A nice
 and detailed introduction to gossip algorithms can be found in
 \cite{shah}.

The model we introduce and analyze in this paper can be viewed as an extension to the
model discussed by Acemoglu et al.~\cite{daron}, who used a gossip algorithm to describe
the spread of misinformation in social networks. In their model, the state of
each node is viewed as its belief and the randomized gossip
algorithm characterizes  the dynamics of the belief evolution. We
believe that our model is one of the first to consider faulty and
misbehaving nodes in gossip algorithms. While the distributed systems community has
since long recognized the need to provide fault tolerant systems, {e.g.},~\cite{psl:80,dlpsw:86},
efforts to provide similar results for randomized gossiping algorithms
have so far been limited. This paper aims at providing such
results.

\subsection{Main Contribution}
The main contribution of this paper is to provide conditions for
agreement convergence and disagreement divergence over random
networks. To study this problem, we use a model of asynchronous
randomized gossiping.  At each instance, two nodes are selected to
meet with a given probability. When nodes meet, normally they should
update as a weighted average (attraction). Besides that, we assume  that nodes can
misbehave in the sense that they can take a weighted combination with one negative
coefficient (repulsion), or they can stick to their current state
(neglect). The potential node misbehavior essentially results
in model uncertainties in the considered  algorithm. Each node follows
one of the three update rules at random by given probabilities
whenever it is selected to meet another node.
From an engineering viewpoint, this is a simple model of faults or
data attacks for distributed computations. From a
social network perspective, the model constitutes  a non-Bayesian
framework for describing how trust and mistrust of information can propagate in a
society.

A fundamental question we answer is whether the network will
converge to agreement (all nodes asymptotically reach the same value a.s.) or diverge to
disagreement (all nodes disperse a.s.). We study both
symmetric and asymmetric node updates \cite{f-z}. Two general
impossibility theorems are first proposed. Then, a series
of sufficient and/or necessary  conditions are established for the
network to reach a.s. agreement convergence or disagreement divergence. In particular, under symmetric updates, a critical
convergence  measure is found in the sense that the asymptotic evolution of the network states transits
from agreement to disagreement when this measure switches from negative
to positive. This critical measure is in fact independent of the  structure of the underlying communication  graph. In other words, under the node dynamics considered in this paper, there is no difference if the underlying network is an Erd\"{o}s--R\'{e}nyi  graph \cite{er}, a  small-world graph  \cite{watts}, or a scale-free graph \cite{bara}, for the network to reliably target  an agreement.

\subsection{Outline}
The rest of the paper is organized as follows. In Section~2, we
introduce the network model, the considered algorithm, the problem
formulation, together with some physical motivation for the model. Section~3 presents two general
impossibility theorems on a.s. agreement and disagreement, respectively. In Section~4, we
discuss the model in the absence of node repulsion and give conditions
for a.s. agreement convergence for both symmetric and asymmetric
update steps. Section~5 presents agreement and disagreement conditions
for the general model. Finally, some concluding remarks are given in
Section~6.

\section{Problem Definition}
In this section, we present the considered network model and define the problem of interest.

We first  recall  some  basic definitions from graph theory \cite{god} and stochastic matrices \cite{lat}. A directed graph (digraph) $\mathcal
{G}=(\mathcal {V}, \mathcal {E})$ consists of a finite set
$\mathcal{V}$ of nodes and an arc set
$\mathcal {E}\subseteq \mathcal{V}\times\mathcal{V}$.  An element $e=(i,j)\in\mathcal {E}$ is  an
{\it arc}  from node $i\in \mathcal{V}$  to $j\in\mathcal{V}$. A digraph $\mathcal {G}$ is  bidirectional if for every two nodes $i$ and $j$, $(i,j)\in \mathcal{E}$  if and only if $(j,i)\in \mathcal{E}$; $\mathcal
{G}$ is  {\it weakly connected} if it is connected as a bidirectional graph when all the arc directions are ignored. The {\it converse graph}, $\mathcal{G}^T$ of a digraph $\mathcal
{G}=(\mathcal {V}, \mathcal {E})$, is defined as the graph obtained by reversing the orientation of all arcs in $ \mathcal {E}$. A finite square matrix $M=[m_{ij}]\in\mathds{R}^{n\times n}$ is called {\em stochastic} if $m_{ij}\geq 0$ for all $i,j$ and $\sum_j m_{ij}=1$ for all $i$.  A stochastic matrix $M$ is  {\em doubly stochastic} if also $M^T$ is  stochastic. Let $P=[p_{ij}]\in\mathds{R}^{n\times n}$ be a matrix with nonnegative entries. We can associate a unique digraph  $\mathcal{G}_P=(\mathcal{V},\mathcal{E}_P)$ with $P$ on node set $\mathcal{V}=\{1,\dots,n\}$ such that $(j,i)\in\mathcal{E}_P$ if and only if $p_{ij}>0$. We call $\mathcal{G}_P$ the {\em induced graph} of $P$.

\subsection{Node Pair Selection}

Consider a network  with node set $\mathcal{V}=\{1,\dots,n\}$, $n\geq3$.  Let the digraph $\mathcal {G}_0=(\mathcal{V},\mathcal{E}_0)$ denote the {\it underlying}  graph of the considered network. The underlying graph indicates  potential interactions between nodes. We use the asynchronous time model introduced in \cite{boyd} to describe node  interactions. Each node meets other nodes at independent time instances  defined  by a  rate-one Poisson process. This is to say, the inter-meeting times at each node follows a rate-one  exponential distribution. Without loss of generality, we can assume that  at most one node is active at any given instance. Let $x_i(k)\in\mathds{R}$ denote the state (value) of node $i$ at the $k$'th meeting slot among all the nodes.

  Node interactions are characterized by an $n\times n$ matrix $A=[a_{ij}]$, where $a_{ij}\geq0$ for all $i,j=1,\dots,n$ and $a_{ij}>0$ if and only if $(j,i)\in \mathcal{E}_0$. We assume $A$ is a stochastic matrix. The meeting process is defined as follows.

\begin{defn}[Node Pair Selection] Independent of time and node state, at time $k\geq0$,
\begin{itemize}
\item[(i)]  A node $i\in\mathcal{V}$ is drawn    with probability $1/n$;
  \item[(ii)] Node $i$ picks the pair $(i,j)$ with probability $a_{ij}$.
\end{itemize}
\end{defn}

Note that, by the definition of the node pair selection process, the underlying graph $\mathcal{G}_0$ is actually the same as $\mathcal{G}_A$, the induced graph of the node pair selection matrix $A$.  For $\mathcal{G}_0$, we use the following assumption.

\vspace{3mm}

  \noindent {\bf A1.} {\em (Underlying Connectivity)} The underlying graph $\mathcal {G}_0$ is weakly connected.

\subsection{State Evolution}
Suppose  node $i$ meets another node $j$  at time $k$. Independent of time, node states, and pair selection process, their will be three events for the iterative update for node $i$.
\begin{itemize}
\item[(i)] {\it (Attraction)} With probability $\alpha$, node $i$ updates as a weighted average with $j$, marked by event $\mathscr{A}_{ij}(k)$:
\begin{align}\label{trust}
x_i(k+1)=x_i(k)+ T_k \big(x_j(k)-x_i(k)\big)=(1- T_k)x_i(k)+ T_k x_j(k),
\end{align}
where $0<T_k\leq 1$ is the average weight.

\item[(ii)] {\it (Neglect)}  With probability $\beta$, node $i$ sticks to its current state, marked by event $\mathscr{N}_{ij}(k)$:
\begin{align}
x_i(k+1)=x_i(k).
\end{align}

\item[(iii)] {\it (Repulsion)} With probability $\gamma$, node $i$  updates as a weighted average with $j$, but with a negative coefficient, marked by  $\mathscr{R}_{ij}(k)$:
\begin{align}
x_i(k+1)=x_i(k)- S_k\big(x_j(k)-x_i(k)\big)=(1+S_k)x_i(k)- S_kx_j(k),
\end{align}
where $S_k>0$.
\end{itemize}
Naturally we  assume $\alpha+\beta+\gamma=1$. Node $j$'s update is determined by the corresponding events $\mathscr{A}_{ji}(k)$, $\mathscr{N}_{ji}(k)$ and $\mathscr{R}_{ji}(k)$, which may depend on  node $i$'s update.

\subsection{Problem}

Let  $x^0=x(k_0)=(x_1(k_0)\dots x_n(k_0))^T\in \mathds{R}^{n}$ be the initial condition, where $k_0\geq 0$ is an arbitrary integer. Denote $
x(k;k_0,x^0)=\big(x_1\big(k;k_0,x_1(k_0)\big)\dots x_n\big(k;k_0,x_n(k_0)\big)\big)^T\in \mathds{R}^{n}$ as the random process driven by the randomized update. When it is clear from the context, we will identify $x(k;k_0,x^0)$ with $x(k)$.

Introduce
$$
 H(k)\doteq\max_{i\in\mathcal{V}}x_{i}(k), \quad h(k)\doteq\min_{i\in\mathcal{V}}x_{i}(k)
$$
as the maximum and minimum  states among all nodes, respectively, and define $\mathcal{H}(k)\doteq H(k)-h(k)$ as the agreement measure. We make the following definition.
\begin{defn}[Agreement Convergence and Disagreement Divergence]\ \ \

\begin{itemize}
\item[(i)] Agreement convergence is achieved  a.s. for initial value $x(k_0)\in \mathds{R}^{n}$  if
\begin{equation}\label{100}
\mathbf{P}\Big(\limsup_{k\rightarrow \infty} \mathcal{H}(k)=0\Big)=1.
\end{equation}
Global agreement convergence is achieved a.s. if (\ref{100}) holds for all initial values.

\item[(ii)] Disagreement divergence is achieved a.s.  for initial value $x(k_0)\in \mathds{R}^{n}$ if
\begin{equation}
\mathbf{P}\Big(\limsup_{k\rightarrow \infty} \mathcal{H}(k)>M\Big)=1\  for\ all\ M\geq 0.
\end{equation}
\end{itemize}
\end{defn}

Agreement convergence corresponds to that all states asymptotically reach the same value. Disagreement divergence does not only mean that agreement is not achieved, but that the difference of the maximum and minimum states asymptotically diverges.
\subsection{Model Rationale}
We illustrate and motivate the model introduced above through three
application examples.

\subsubsection*{False Data Injection Attacks}
Large distributed computing and control systems are vulnerable to
cyber attacks. An attacker may inject false data or malicious
code in the network, to mislead the nodes or even change the overall behavior of
the system. The model in this paper can represent a very simple system
under a cyber attack. The attraction event $\mathscr{A}_{ij}$ corresponds to normal
operation of the system, under which the nodes are supposed to reach
consensus. The neglect event $\mathscr{N}_{ij}$ can represent
a denial-of-service attack, which block node $i$ from updating
its state based on information from its neighbor $j$. The injection of
malicious code in node $i$ changing its  update law is modeled by
the repulsion update. State agreement or disagreement indicates the
failure or success of the attack. Our results in this paper allow us
to explicitly characterize how  large attacks a network
can withstand.
Various false data injection attacks for dynamical systems have recently been discussed in
the literature, e.g., \cite{attack1,attack2,attack3,attack4}.

\subsubsection*{Fault-Tolerant Systems}
 ``An important goal in distributed system design is to construct the system in such a way that it
can automatically recover from partial failures without seriously affecting the overall
performance,'' as pointed out in \cite{ft1}. In our model the events
$\mathscr{N}_{ij}$ and $\mathscr{R}_{ij}$ can represent node
faults during a randomized computation process or in the coordination
of a multi-robot system. For example, the
magnitude of the repulsion parameter $S_k$ can indicate how severe a
fault is. Our results show that a networked systems can sometimes be
robust to quite severe faults. It is also shown that in certain cases
the topology of the network does not play an essential role but the
persistence and the size are more important.

\subsubsection*{Social Networks}
Started from the classical work of DeGroot~\cite{social1}, distributed
averaging similar to our model has been widely used to characterize opinion dynamics in
social networks, e.g., \cite{social2,social3,daron,como}. The state
$x_i$ of node $i$ is in these models the opinion of an
individual. The individuals meet and  exchange opinions. The
attraction event $\mathscr{A}_{ij}$ models the trust of node $i$ to
node $j$. Whenever $\mathscr{A}_{ij}(k)$ happens, node $i$ believes in
node $j$ and therefore takes an attraction update step. The parameter
$T_k$ measures the level of trust. The neglect event
$\mathscr{N}_{ij}$ models the mistrust of node $i$ to node $j$, which
results in that $i$ simply ignores $j$ and sticks to its current
opinion. The repulsion event $\mathscr{R}_{ij}$ models the antagonism
of node $i$ to node $j$. In this case, node $i$ takes the opposite
direction to the attraction to keep a large distance to the opinion of
node $j$. In this way, our model characterizes the influence of node
relations to the convergence of the opinion in social networks. The
idea follows the discussions on the possibilities of spread of
misinformation and persistent disagreement in \cite{daron,como}. In
addition, our model also allows for opinion divergence, as
indicated in the definition of disagreement divergence.

\section{Impossibility Theorems}
In this section, we discuss the impossibilities  of agreement convergence or disagreement divergence.

 A general impossibility theorem  for agreement convergence is established  as follows on the sequence $\{T_k\}_0^\infty$, i.e., on the strength of the attraction step (\ref{trust}).

\begin{thm}\label{thm1}
Global agreement convergence  can be achieved a.s.  only if either $\sum_{k=0}^\infty T_k=\infty$ or $\sum_{k=0}^\infty (1- T_k)=\infty$. In fact, if either $\sum_{k=0}^\infty T_k<\infty$ or $\sum_{k=0}^\infty (1- T_k)<\infty$ holds, then for almost all initial values, we have
\begin{equation}
\mathbf{P}\Big(\limsup_{k\rightarrow \infty} \mathcal{H}(k)=0\Big)=0
\end{equation}
when $k_0$ is sufficiently large.
\end{thm}
{\it Proof.} The proof relies on the following  well-known lemma.
\begin{lem}\label{lem1}
Let $\{b_k\}_0^\infty$ be a sequence of real numbers with $b_k\in[0,1)$ for all $k$. Then $\sum_{k=0}^\infty b_k=\infty$ if and only if $\prod_{k=0}^{\infty}(1-b_k)=0$.
\end{lem}

Now suppose $\sum_{k=0}^\infty T_k<\infty$. Then $\exists K_0 \geq 0$ s.t. $T_k<1/2, k\geq K_0$. Let node pair $(i,j)$ be selected at time $k\geq K_0$. There are two cases.
\begin{itemize}
\item[(i)] Neither $x_i(k)$ nor $x_j(k)$ reaches the minimum value. Then straightforwardly we have
$$h(k+1)\leq h(k).$$
\item[(ii)] One of the two nodes, say $i$, reaches the minimum value. In this case, we have
\begin{align}
x_i(k+1)=x_i(k)+ T_k \big(x_j(k)-x_i(k)\big)\leq h(k)+ T_k \mathcal{H}(k) \nonumber
\end{align}
 if $\mathscr{A}_{ij}(k)$ happens,  and  $h(k+1)\leq h(k)$ otherwise.
\end{itemize}
Thus, we obtain
\begin{align}\label{s1}
\mathbf{P}\Big( h(k+1)\leq h(k)+ T_k \mathcal{H}(k),\ k\geq K_0\Big)=1.
\end{align}

A similar analysis leads to that
\begin{align}\label{s2}
\mathbf{P}\Big( H(k+1)\geq H(k)- T_k \mathcal{H}(k),\ k\geq K_0\Big)=1.
\end{align}
We see from (\ref{s1}) and (\ref{s2}) that
\begin{align}\label{s3}
\mathbf{P}\Big( \mathcal{H}(k+1)\geq \big(1- 2T_k\big) \mathcal{H}(k),\ k\geq K_0\Big)=1.
\end{align}

Thus, according to (\ref{s3}), we conclude
\begin{align}
\mathbf{P}\Big( \mathcal{H}(m)\geq \prod_{k=K_0}^ m\big(1- 2T_k\big) \mathcal{H}(K_0) \geq  \prod_{k=K_0}^ \infty\big(1- 2T_k\big)\mathcal{H}(K_0)= \rho_\ast \mathcal{H}(K_0) \Big)=1 \nonumber
\end{align}
for all $m\geq K_0$, where $\rho_\ast\doteq \prod_{k=K_0}^ \infty\big(1- 2T_k\big)$ is a constant satisfying $0<\rho_\ast<1$ based on Lemma \ref{lem1}. This implies
\begin{align}
\mathbf{P}\Big(\limsup_{k\rightarrow \infty} \mathcal{H}(k)>0\Big)\geq \mathbf{P}\Big( \mathcal{H}(m)>0,\ m\geq K_0 \Big)=1 \nonumber
\end{align}
for all initial conditions with $k_0\geq K_0$ and $\mathcal{H}(k_0)>0$. It is obvious to see that
$$
\big\{x=(x_1\dots x_n)^T \in \mathds{R}^n: x_1=\dots=x_n\big\}
$$ is a set with measure zero in $\mathds{R}^n$. The desired conclusion follows.

Moreover,  the conclusion  for the other case  $\sum_{k=0}^\infty (1- T_k)<\infty$ follows from a symmetric argument. This completes the proof.
\hfill$\square$

Theorem \ref{thm1} establishes two general lower bounds for  the attraction update  regarding    a.s. agreement convergence. Note that Theorem \ref{thm1} does not impose any assumption on the dependence of $\mathscr{A}_{ij}(k)$ and $\mathscr{A}_{ji}(k)$ for the conclusion to stand.

The corresponding impossibility theorem for disagreement divergence is presented as follows.
 \begin{thm}\label{thm100}
Disagreement divergence  can be achieved a.s.  only if $\prod_{k=0}^\infty (1+2S_k)=\infty$.
\end{thm}
{\it Proof.} According to the definition of the randomized dynamics, it is straightforward to see that
\begin{align}
\mathbf{P}\Big( \mathcal{H}(k+1)\leq \big(1+ 2S_k\big) \mathcal{H}(k) \Big)=1
\end{align}
for all $k$. The desired conclusion follows immediately. \hfill$\square$

In the rest of the paper, we turn to the possibilities of a.s. agreement convergence and disagreement divergence.

\section{Attraction  vs. Neglect}
In this section, we focus on the role of node attraction  for the network to reach a.s. agreement convergence.  In the absence of node repulsion, we show how much attraction update is enough to guarantee global a.s. agreement under symmetric or asymmetric node dynamics, respectively.

\subsection{Symmetric Update}
Consider the case when  repulsion events never take place, i.e., nodes can only follow the attraction or neglect events.  We use the following assumption, where by definition a trivial event has probability zero.

\vspace{2mm}
\noindent {\bf A2.} {\it (Repulsion-Free)} $\mathscr{R}_{ij}(k)$ is a trivial event for all $(i,j)$ and $k$.
\vspace{2mm}

This subsection focuses on the condition when the nodes' updates are symmetric  when two nodes meet, as indicated in the following assumption.

\vspace{2mm}
\noindent {\bf A3.} {\em (Symmetric Attraction)} The events $\mathscr{A}_{ij}(k)=\mathscr{A}_{ji}(k)$ for all $(i,j)$ and $k$.

The main result for the symmetric  update model is as follows.
\begin{prop}\label{thm2}
Suppose {\bf A1},  {\bf A2}  and {\bf A3} hold. Global agreement convergence  is achieved a.s. if $\sum_{k=0}^\infty T_k(1-T_k)=\infty$.
\end{prop}
{\it Proof.} With {\bf A2}  and {\bf A3}, the considered gossip algorithm can be expressed as
 \begin{align}\label{1}
 x(k+1)=\Phi(k)x(k),
 \end{align}
where $\Phi(k)$ is the random matrix satisfying
\begin{align}\label{2}
\mathbf{P}\Big(\Phi(k)=\Phi_{\langle ij\rangle}\doteq I-T_k (e_i-e_j)(e_i-e_j)^T\Big)=\frac{\alpha}{n}(a_{ij}+a_{ji}),\ \ \ \  i\neq j
\end{align}
 with $e_m=(0 \dots 0\  1\  0 \dots 0)^T$ denoting the $n\times1$ unit vector whose $m$'th component is $1$.

 Define $L(k)=\sum_{i=1}^n |x_i(k)-x_{\rm ave}|^2$, where $x_{\rm ave}=\sum_{i=1}^n{x_i(k_0)}/n$ is the average of the initial values and $|\cdot|$ represents the Euclidean norm of a vector or the absolute value of a scalar.

It is easy to verify  that for every possible sample and fixed instant $k$, $\Phi_{\langle ij\rangle}$  defined in (\ref{2}),  is a  symmetric, and doubly stochastic matrix, i.e., $\Phi_{\langle ij\rangle}\mathbf{1}=\mathbf{1}$ and $\mathbf{1}^T \Phi_{\langle ij\rangle}=\mathbf{1}^T$.

Therefore, we have
\begin{align}\label{3}
\mathbf{E}\Big(L(k+1)\big|x(k)\Big)&=\mathbf{E}\Big( \big(x(k+1)-x_{\rm ave}\mathbf{1}\big)^T \big(x(k+1)-x_{\rm ave}\mathbf{1}\big)\big|x(k)\Big)\nonumber\\
&=\mathbf{E}\Big( \big(\Phi(k)x(k)-x_{\rm ave}\mathbf{1}\big)^T \big(\Phi(k)x(k)-x_{\rm ave}\mathbf{1}\big)\big|x(k)\Big)\nonumber\\
&=\mathbf{E}\Big( \big(x(k)-x_{\rm ave}\mathbf{1}\big)^T\Phi(k)^T\Phi(k) \big(x(k)-x_{\rm ave}\mathbf{1}\big)\big|x(k)\Big)\nonumber\\
&=\big(x(k)-x_{\rm ave}\mathbf{1}\big)^T \mathbf{E}\big(\Phi^2(k)\big) \big(x(k)-x_{\rm ave}\mathbf{1}\big)
\end{align}

Since  every possible sample of $\Phi(k)$ is doubly stochastic, each sample of $\Phi^2(k)$ is also doubly stochastic. This implies  $\mathbf{E}\big(\Phi(k)^T\Phi(k)\big)$ is a stochastic matrix for all $k$, and $\mathbf{1}$ is the eigenvector corresponding to eigenvalue $1$ of $\mathbf{E}\big(\Phi^2(k)\big)$. Thus, we can conclude from (\ref{3}) that
\begin{align}\label{6}
\mathbf{E}\Big(L(k+1)\big|x(k)\Big)&\leq \lambda_2 \Big(\mathbf{E}\big(\Phi^2(k)\big)\Big)\big(x(k)-x_{\rm ave}\mathbf{1}\big)^T  \big(x(k)-x_{\rm ave}\mathbf{1}\big)\nonumber\\
&=\lambda_2 \Big(\mathbf{E}\big(\Phi^2(k)\big)\Big)L(k),
\end{align}
where $\lambda_2(M)$ for a stochastic matrix $M$ denotes the largest eigenvalue in magnitude excluding the eigenvalue at one.

Noticing that
\begin{align}\label{20}
\Big(I-T_k (e_i-e_j)(e_i-e_j)^T\Big)^2=I- 2T_k(1-T_k)(e_i-e_j)(e_i-e_j)^T
\end{align}
 we see from (\ref{2}) that
\begin{align}
\mathbf{P}\Big(\Phi^2(k)= I- 2T_k(1-T_k)(e_i-e_j)(e_i-e_j)^T\Big)=\frac{\alpha}{n}(a_{ij}+a_{ji}),\ \ \ \  i\neq j. \nonumber
\end{align}
This leads to
\begin{align}\label{4}
\mathbf{E}\big(\Phi^2(k)\big)= I-2T_k(1-T_k) \frac{\alpha}{n}\big(D-(A+A^T)\big),
\end{align}
where  $D=\mbox{diag}(d_1 \dots d_n)$ with $d_i=\sum_{j=1}^n (a_{ij}+a_{ji})$.

 Note that $D-(A+A^T)$ is actually the (weighted) Laplacian of the graph $\mathcal{G}_{A+A^T}$. With assumption {\bf A1}, $\mathcal{G}_{A+A^T}$ is a connected graph,  and therefore, based on the well-known property of Laplacian matrix of connected graphs \cite{god}, we have $\lambda_2^\ast>0$, where $\lambda_2^\ast$ is the second smallest eigenvalue of $D-(A+A^T)$. On the other hand, since $A$ is a stochastic matrix, it is straightforward to see that
 \begin{align}
 \sum_{j=1,j\neq i}a_{ij}+a_{ji}\leq n
 \end{align}
for all $i=1,\dots,n$. According to Gershgorin's circle theorem, every  eigenvalue $\lambda_i^\ast$ of $D-(A+A^T)$ is bounded by $2n$. Therefore,
\begin{align}\label{5}
0<2T_k(1-T_k) \frac{\alpha}{n} \lambda_i^\ast \leq 4 T_k(1-T_k) \leq 4\Big(\frac{T_k+(1-T_k)}{2}\Big)^2=1
\end{align}
for all $\lambda_i^\ast\in \sigma\big(D-(A+A^T)\big)$, where $\sigma(\cdot)$ denotes the spectrum of a matrix.

Now we conclude from (\ref{4}) and (\ref{5}) that for all $k$,
\begin{align}\label{7}
\lambda_2 \Big(\mathbf{E}\big(\Phi^2(k)\big)\Big)= 1- \frac{2T_k(1-T_k)\alpha}{n}\lambda_2^\ast.
\end{align}
With (\ref{6}) and (\ref{7}), we obtain
\begin{align}\label{10}
\mathbf{E}\Big(L(k+1)\Big)
\leq \prod_{i=k_0}^{k}\lambda_2 \Big(\mathbf{E}\big(\Phi^2(i)\big)\Big)L(k_0)=\prod_{i=k_0}^{k}\Big(1-\frac{2T_k(1-T_k)\alpha}{n}\lambda_2^\ast\Big)L(k_0),
\end{align}

Therefore, based on Lemma \ref{lem1} and Fatou's lemma, we have
\begin{align}
\mathbf{E}\Big(\lim_{k\rightarrow\infty}L(k)\Big)\leq\lim_{k\rightarrow\infty}\mathbf{E}\Big(L(k)\Big)=0, \nonumber
\end{align}
if $\sum_{k=0}^\infty T_k(1-T_k)=\infty$,
where $\lim_{k\rightarrow\infty}L(k)$ exists simply from the fact that the sequence is non-increasing. This immediately implies
\begin{align}
\mathbf{P}\Big(\lim_{k\rightarrow\infty} x_i(k)=x_{\rm ave}\Big)=1. \nonumber
\end{align}
The proof is finished. \hfill$\square$

There is an interesting connection between the impossibility statement Theorem \ref{thm1} and Proposition \ref{thm2}. Let us consider a special case when $T_k$ is monotone. It is not hard to find that if $T_{k+1}\leq T_k$ for all $k$,  $\sum_{k=0}^\infty T_k(1-T_k)<\infty$ leads to $\sum_{k=0}^\infty T_k<\infty$, while  $\sum_{k=0}^\infty T_k(1-T_k)<\infty$ leads to $\sum_{k=0}^\infty (1-T_k)<\infty$ if $T_{k+1}\geq T_k$ for all $k$. As a result, combining Theorem \ref{thm1} and Proposition \ref{thm2}, we have the following conclusion.

\begin{thm}\label{prop1}
Suppose {\bf A1}, {\bf A2} and {\bf A3} hold. Assume that either $T_{k+1}\leq T_k$ or $T_{k+1}\geq T_k$ for all $k$. Then  $\sum_{k=0}^\infty T_k(1-T_k)=\infty$ is a threshold condition regarding global a.s. agreement convergence:

(i) $\mathbf{P}\big(\limsup_{k\rightarrow \infty} \mathcal{H}(k)=0\big)=0$ for almost all initial conditions with $k_0$ sufficiently large if $\sum_{k=0}^\infty T_k(1-T_k)<\infty$;

(ii) $\mathbf{P}\big(\limsup_{k\rightarrow \infty} \mathcal{H}(k)=0\big)=1$ for all initial conditions if $\sum_{k=0}^\infty T_k(1-T_k)=\infty$.
\end{thm}

\subsection{Asymmetric  Update}
In this subsection, we investigate the case when the node updates are asymmetric, as indicated by the following assumption.

\vspace{2mm}
\noindent {\bf A4.} {\em (Asymmetric Attraction)}  $\mathscr{A}_{ij}(k)\bigcap \mathscr{A}_{ji}(k)$ is a trivial event for all $(i,j)$ and $k$.

 We present the main result for the asymmetric  update  model  as follows.
\begin{prop}\label{thm3}
Suppose {\bf A1}, {\bf A2} and {\bf A4} hold. Then global agreement convergence  is achieved a.s. if
\begin{align}
\sum_{k=0}^\infty \bigg[ \prod_{s=k(n-1)}^{ (k+1)(n-1)-1} T_s\big(1-T_s\big)\bigg]=\infty. \nonumber
\end{align}
\end{prop}
{\it Proof.} Take $k_\ast\geq 0$. Denote $a_\ast=\min\{a_{ij}:\ a_{ij}>0\}$ as the lower bound of the nonzero entries of $A$. Suppose $i_0$ is some node satisfying $x_{i_0}(k_\ast)=h(k_\ast)$.

 Let $i_1$  be a node which is connected to $i_0$ in graph  $\mathcal{G}_0^\ast$.  We see that  such $i_1$ exists based on the weak connectivity assumption {\bf A1}.   With assumptions {\bf A2} and {\bf A4}, we have
\begin{align}
\mathbf{P}\Big(\mbox{pair $(i_0,i_1)$  or $(i_1,i_0)$ selected, and $\mathscr{A}_{i_1i_0}$ happens} \Big)\geq \frac{a_\ast }{n}\alpha. \nonumber
\end{align}
Moreover, if $\mathscr{A}_{i_1i_0}$ happens, we have
\begin{align}
x_{i_1}(k_\ast+1)&=x_{i_1}(k_\ast)+T_{k_\ast}\big(x_{i_0}(k_\ast)-x_{i_1}(k_\ast)\big)\nonumber\\
&= T_{k_\ast}x_{i_0}(k_\ast)+(1- T_{k_\ast})x_{i_1}(k_\ast)\nonumber\\
&\leq T_{k_\ast}h(k_\ast)+(1- T_{k_\ast})H(k_\ast)\nonumber\\
&\leq T_{k_\ast}(1-T_{k_\ast})h(k_\ast)+\big(1- T_{k_\ast}(1-T_{k_\ast})\big)H(k_\ast) \nonumber
\end{align}
and $x_{i_0}(k_\ast+1)=x_{i_0}(k_\ast)$ according to assumption  {\bf A4}. This implies
\begin{align}
\mathbf{P}\Big(x_{i_1}(k_\ast+1) \leq  T_{k_\ast}(1-T_{k_\ast})h(k_\ast)+\big(1- T_{k_\ast}(1-T_{k_\ast})\big)H(k_\ast)\ \mbox{and}\ x_{i_0}(k_\ast+1)=x_{i_0}(k_\ast)  \Big)\geq \frac{a_\ast }{n}\alpha. \nonumber
\end{align}

\vspace{2mm}

Next, according to  the weak connectivity assumption {\bf A1}, another node $i_2$ can be found such that $i_2$ is connected to $\{i_0,i_1\}$ in  $\mathcal{G}_0^\ast$. There will be two cases.
\begin{itemize}
\item[(i)] $i_2$ is connected to $i_0$ in  $\mathcal{G}_0^\ast$. Then by a similar analysis we used for bounding  $x_{i_1}(k_\ast+1)$, we obtain
\begin{align}
&\mathbf{P}\Big(x_{i_0}(k_\ast+2)=x_{i_0}(k_\ast),\ x_{i_1}(k_\ast+2)=x_{i_1}(k_\ast+1),\nonumber\\
 &\ \ \ \ \ \ \ \ \ \ \ \ \mbox{and}\ x_{i_2}(k_\ast+2) \leq T_{k_\ast+1}h(k_\ast)+(1- T_{k_\ast+1})H(k_\ast)\ \  \Big)\geq \frac{ a_\ast }{n}\alpha. \nonumber
\end{align}
\item[(ii)] $i_2$ is connected to $i_1$ in  $\mathcal{G}_0^\ast$. Suppose pair $(i_1,i_2)$  or $(i_2,i_1)$ selected, and $\mathscr{A}_{i_2i_1}$ happens at time $k_\ast+1$. Then we have
    \begin{align}
    x_{i_1}(k_\ast+2)=x_{i_1}(k_\ast+1) \nonumber
    \end{align}
    and
\begin{align}
x_{i_2}(k_\ast+2)&=x_{i_2}(k_\ast+1)+T_{k_\ast+1}\big(x_{i_1}(k_\ast)-x_{i_2}(k_\ast+1)\big)\nonumber\\
&\leq (1-T_{k_\ast+1})H(k_\ast+1)+T_{k_\ast+1}\Big( T_{k_\ast}(1-T_{k_\ast})h(k_\ast)+\big(1- T_{k_\ast}(1-T_{k_\ast})\big)H(k_\ast)\Big) \nonumber\\
&\leq  (1-T_{k_\ast+1})H(k_\ast)+T_{k_\ast+1}\Big( T_{k_\ast}(1-T_{k_\ast})h(k_\ast)+\big(1- T_{k_\ast}(1-T_{k_\ast})\big)H(k_\ast)\Big)\nonumber\\
&=T_{k_\ast+1} T_{k_\ast}(1-T_{k_\ast})h(k_\ast)+\Big(1-T_{k_\ast+1} T_{k_\ast}(1-T_{k_\ast})\Big)H(k_\ast)\nonumber\\
&\leq h(k_\ast)\prod_{k=k_\ast}^{k_\ast+1} T_{k}(1-T_{k})+H(k_\ast)\Big(1-\prod_{k=k_\ast}^{k_\ast+1} T_{k}(1-T_{k})\Big) \nonumber
\end{align}
conditioned  that pair $(i_0,i_1)$  or $(i_1,i_0)$ selected, and $\mathscr{A}_{i_1i_0}$ happens at time $k_\ast$.
\end{itemize}

Noting the fact that
$$
T_{k_\ast}h(k_\ast)+(1- T_{k_\ast})H(k_\ast)\leq h(k_\ast)\prod_{k=k_\ast}^{k_\ast+1} T_{k}(1-T_{k})+H(k_\ast)\Big(1-\prod_{k=k_\ast}^{k_\ast+1} T_{k}(1-T_{k})\Big),
$$
we conclude from either of the two cases that
\begin{align}
&\mathbf{P}\Big(x_{\tau}(k_\ast+2) \leq  h(k_\ast)\prod_{k=k_\ast}^{k_\ast+1} T_{k}(1-T_{k})+H(k_\ast)\big(1-\prod_{k=k_\ast}^{k_\ast+1} T_{k}(1-T_{k})\big),\ \tau=i_0,i_1,i_2\Big)\geq \Big(\frac{\alpha a_\ast }{n}\Big)^2.\nonumber
\end{align}

\vspace{2mm}

Continuing we obtain that for nodes $i_3,\dots,i_{n-1}$,
\begin{align}
&\mathbf{P}\Big(x_{\tau}(k_\ast+n-1) \leq h(k_\ast)\prod_{k=k_\ast}^{k_\ast+n-2}T_{k}(1-T_{k})+H(k_\ast)\Big(1-\prod_{k=k_\ast}^{k_\ast+n-2}T_{k}(1-T_{k})\Big),\nonumber\\ &\quad \quad \quad \quad \quad \quad \quad \quad \quad \quad \quad  \tau=i_0,\dots,i_{n-1}\Big)\geq \Big(\frac{\alpha a_\ast }{n}\Big)^{n-1}, \nonumber
\end{align}
which yields
\begin{align}\label{11}
\mathbf{P}\Big(H(k_\ast+n-1) \leq h(k_\ast)\prod_{k=k_\ast}^{k_\ast+n-2}T_{k}(1-T_{k})+H(k_\ast)\Big(1-\prod_{k=k_\ast}^{k_\ast+n-2}T_{k}(1-T_{k})\Big)\Big)\geq \Big(\frac{\alpha a_\ast }{n}\Big)^{n-1}.
\end{align}

With (\ref{11}), we obtain
\begin{align}\label{12}
\mathbf{P}\Big(\mathcal{H}(k_\ast+n-1) \leq \Big(1-\prod_{k=k_\ast}^{k_\ast+n-2}T_{k}(1-T_{k})\Big)\mathcal{H}(k_\ast)\Big)\geq \Big(\frac{\alpha a_\ast }{n}\Big)^{n-1}.
\end{align}
Since assumption {\bf A2} guarantees $\mathcal{H}(k+1)\leq \mathcal{H}(k)$ for all $k$ with probability one, (\ref{12}) leads to
\begin{align}
\mathbf{E}\big(\mathcal{H}(k_\ast+n-1)\big) \leq \Big(1- \Big(\frac{\alpha a_\ast }{n}\Big)^{n-1}\prod_{k=k_\ast}^{k_\ast+n-2}T_k(1-T_{k})\Big)\mathbf{E}\big(\mathcal{H}(k_\ast)\big).
\end{align}

Note that $k_\ast$ is chosen arbitrarily in the upper analysis. We conclude by induction that
\begin{align}
\mathbf{E}\Big(\mathcal{H}\big(k_\ast+(s+1)(n-1)\big)\Big) \leq \Big(1- \Big(\frac{\alpha a_\ast }{n}\Big)^{n-1} \prod_{k=k_\ast+s(n-1)}^{k_\ast+(s+1)(n-1)-1} T_{k} (1-T_k)\Big)\mathbf{E}\Big(\mathcal{H}\big(k_\ast+s(n-1)\big)\Big) \nonumber
\end{align}
for all $s=0,1,\dots$. Particularly, we choose $k_\ast=K_0(n-1)\geq k_0$ for some integer $K_0\geq 0$, where $k_0$ is the initial time, we obtain
\begin{align}
\mathbf{E}\Big(\mathcal{H}\big((s+1)(n-1)\big)\Big) \leq \prod_{t=K_0}^{s}\Big(1- \Big(\frac{\alpha a_\ast }{n}\Big)^{n-1} \prod_{k=t(n-1)}^{(t+1)(n-1)-1} T_{k} (1-T_k) \Big)\mathbf{E}\Big(\mathcal{H}\big(K_0(n-1)\big)\Big), \nonumber
\end{align}
which implies
\begin{align}\label{16}
\mathbf{E}\Big(\lim_{s\rightarrow \infty}\mathcal{H}\big(s(n-1)\big)\Big)\leq \lim_{s\rightarrow \infty}\mathbf{E}\Big(\mathcal{H}\big(s(n-1)\big)\Big)=0
\end{align}
by Fatou's Lemma and Lemma \ref{lem1} as long as  $\sum_{k=0}^\infty  \prod_{s=k(n-1)}^{ (k+1)(n-1)-1} T_s\big(1-T_s\big)=\infty$. Therefore, observing that $\mathcal{H}(k)$ is non-increasing, (\ref{16}) leads to
\begin{equation}
\mathbf{P}\Big(\limsup_{k\rightarrow \infty} \mathcal{H}(k)=0\Big)=1.
\end{equation}
The desired conclusion follows.
\hfill$\square$

We see from Propositions \ref{thm2} and \ref{thm3} that it is easier to achieve agreement convergence with symmetric updates, which is consistent with the literature \cite{f-z}.

Again let us consider the case when $T_k$ is monotone. The following lemma holds.
\begin{lem}\label{lem2} Let $\{b_k\}_0^\infty$ be a sequence of real numbers with $b_k\in[0,1]$ for all $k$.

\begin{itemize}
\item[(i)] Suppose $b_{k+1}\leq b_k$ for all $k$. Then the following statements are equivalent.
  \begin{itemize}
  \item[a)] $\sum_{k=0}^\infty \prod_{s=k(n-1)}^{ (k+1)(n-1)-1} b_s(1-b_s)=\infty$;
  \item[b)] $\sum_{s=0}^\infty  \Big(b_s(1-b_s)\Big)^{n-1}=\infty$;
  \item[c)]$\sum_{s=0}^\infty  b_s^{n-1}=\infty$.
\end{itemize}
\item[(ii)] Suppose $b_{k+1}\geq b_k$ for all $k$. Then the following statements are equivalent.
\begin{itemize}
  \item[a)] $\sum_{k=0}^\infty  \prod_{s=k(n-1)}^{ (k+1)(n-1)-1} b_s(1-b_s)=\infty$;
  \item[b)] $\sum_{s=0}^\infty  \Big(b_s(1-b_s)\Big)^{n-1}=\infty$;
  \item[c)]$\sum_{s=0}^\infty  \big(1-b_s\big)^{n-1}=\infty$.
\end{itemize}
\end{itemize}
\end{lem}
{\it Proof.} We just prove (i). Case (ii) holds by a similar analysis. Note that, we have
\begin{align}\label{30}
 (1-b_0)^{n-1}b_{(k+1)(n-1)}^{n-1}\leq\prod_{s=k(n-1)}^{ (k+1)(n-1)-1} b_s(1-b_s)\leq  \prod_{s=k(n-1)}^{ (k+1)(n-1)-1} b_s \leq b_{k(n-1)}^{n-1},
\end{align}
where without loss of generality we assume $b_0>0$.  Moreover, the monotonicity of $\{b_k\}$ guarantees
\begin{align}\label{31}
(n-1)\sum_{k=0}^\infty b_{(k+1)(n-1)}^{n-1}\leq\sum_{k=0}^\infty  b_k^{n-1} \leq (n-1)\sum_{k=0}^\infty b_{k(n-1)}^{n-1}.
\end{align}
We see from (\ref{30}) and (\ref{31}) that statements a) and c) are equivalent.

On the other hand, observing that
\begin{align}
(1-b_0)^{n-1} \sum_{s=0}^\infty  b_s^{n-1} \leq \sum_{s=0}^\infty  \Big(b_s(1-b_s)\Big)^{n-1}\leq \sum_{s=0}^\infty  b_s^{n-1}, \nonumber
\end{align}
also statements b) and c) are equivalent. The desired conclusion follows. \hfill$\square$

Combining Proposition  \ref{thm3} and Lemma \ref{lem2}, we obtain the following conclusion.
\begin{thm}\label{prop2}
Suppose {\bf A1}, {\bf A2} and {\bf A3} hold. Assume that either $T_{k+1}\leq T_k$ or $T_{k+1}\geq T_k$ for all $k$. Then
global agreement convergence is achieved a.s. if
\begin{align}
\sum_{k=0}^\infty  \Big(\big(1-T_k\big)T_k\Big)^{n-1}=\infty. \nonumber
\end{align}
\end{thm}

We see from Theorems \ref{prop1} and \ref{prop2} that the requirement for the  sequence $\{T_k\}_0^\infty$ to guarantee a.s.   agreement  convergence increases from $\sum_{k=0}^\infty  T_k(1-T_k)=\infty$ to $\sum_{k=0}^\infty  \big((1-T_k)T_k\big)^{n-1}=\infty$ when the update transits from symmetric to asymmetric. Hence, these results quantify the cost of asymptotic updates versus  the strength of attraction.

\section{Attraction  vs. Repulsion}
In this section, we discuss the case when  node repulsion is present in the model. Intuitively the main challenge here is whether the network reaches agreement convergence, or disagreement divergence, depending on which one can beat another one among node attraction and node repulsion  In the following, we study symmetric and asymmetric updates, respectively.

\subsection{Symmetric Update}
We first consider the case when the node updates are symmetric, as described in the following assumption.

\vspace{2mm}
\noindent {\bf A5.} {\em (Symmetric Update)} The events $\mathscr{A}_{ij}(k)= \mathscr{A}_{ji}(k)$ and $\mathscr{R}_{ij}(k)= \mathscr{R}_{ji}(k)$  for all $(i,j)$ and $k$.

\vspace{2mm}

Let $\lambda_2^\ast$ and $\lambda^\ast_n$ be the second smallest and largest eigenvalues of $D-(A+A^T)$ with $D=\mbox{diag}(d_1 \dots d_n)$, $d_i=\sum_{j=1}^n (a_{ij}+a_{ji})$, respectively.   The main result on a.s. agreement convergence under symmetric update is stated as follows.
\begin{prop}\label{thm4}
Suppose {\bf A1}  and {\bf A5} hold. Global  agreement convergence is achieved  a.s.  if
\begin{align}
\prod_{k=0}^\infty \Big(1- \frac{2}{n}\mathcal{I}_k\Big)=0, \nonumber
\end{align}
where
\begin{equation}
	\mathcal{I}_k =
	\begin{cases}
		\Big( T_k(1-T_k)\alpha-S_k(1+S_k)\gamma\Big)\lambda_2^\ast, & \text{if $T_k(1-T_k)\geq S_k(1+S_k)$;}\\
\Big( T_k(1-T_k)\alpha-S_k(1+S_k)\gamma\Big) \lambda_n^\ast, & \text{if $T_k(1-T_k)< S_k(1+S_k)$.}\\
	\end{cases}
\end{equation}
\end{prop}
{\it Proof.}  With assumption  {\bf A5}, the considered  algorithm can be expressed as
 \begin{align}\label{34}
 x(k+1)=\Psi(k)x(k),
 \end{align}
where $\Psi(k)$ is a random matrix satisfying
\begin{align}
\mathbf{P}\Big(\Psi(k)=\Psi_{\langle ij\rangle}^+\doteq I-T_k (e_i-e_j)(e_i-e_j)^T\Big)=\frac{\alpha}{n}(a_{ij}+a_{ji}),\ \ \ \  i\neq j \nonumber
\end{align}
corresponding to event $\mathscr{A}_{ij}(k)$, and
\begin{align}
\mathbf{P}\Big(\Psi(k)=\Psi_{\langle ij\rangle}^-\doteq I+S_k (e_i-e_j)(e_i-e_j)^T\Big)=\frac{\gamma}{n}(a_{ij}+a_{ji}),\ \ \ \  i\neq j.\nonumber
\end{align}
corresponding to event $\mathscr{R}_{ij}(k)$.

Recall that $L(k)=\sum_{i=1}^n |x_i(k)-x_{\rm ave}|^2$, where $x_{\rm ave}=\sum_{i=1}^n{x_i(k_0)}/n$ is the  initial  average. It is crucial to notice that   every possible sample of of the random matrix $\Psi(k)$ is  symmetric and (generalized) stochastic since its row sum  equals one, even though there are negative entries for the matrices $\Psi_{\langle ij\rangle}^-$. Therefore, similar to (\ref{3}), we have
\begin{align}\label{25}
\mathbf{E}\Big(L(k+1)\big|x(k)\Big)= (x(k)-x_{\rm ave}\mathbf{1}\big)^T \mathbf{E}\big(\Psi^2(k)\big) \big(x(k)-x_{\rm ave}\mathbf{1}\big).
\end{align}

Noticing  (\ref{20}) and
\begin{align}
\Big(I+S_k (e_i-e_j)(e_i-e_j)^T\Big)^2=I+2S_k(1+S_k)(e_i-e_j)(e_i-e_j)^T \nonumber
\end{align}
we obtain
\begin{align}\label{21}
\mathbf{E}\big(\Psi^2(k)\big)= I- 2\Big( T_k(1-T_k)\alpha-S_k(1+S_k)\gamma \Big)\frac{1}{n}\Big(D-(A+A^T)\Big).
\end{align}

There are two cases.
\begin{itemize}
\item[(i).] Suppose $T_k(1-T_k)\geq S_k(1+S_k)$. Recalling that every  eigenvalue $\lambda_i^\ast$ of $D-(A+A^T)$ is bounded by $2n$, we have
\begin{align}
0<2\Big( T_k(1-T_k)\alpha-S_k(1+S_k)\gamma \Big)\frac{1}{n} \lambda_i^\ast \leq 2\Big( T_k(1-T_k)\alpha\Big)\frac{1}{n} \lambda_i^\ast\leq  4 T_k(1-T_k) \leq 1 \nonumber
\end{align}
for all $\lambda_i^\ast\in \sigma\big(D-(A+A^T)\big)$. Thus, all the eigenvalues of $\mathbf{E}\big(\Psi^2(k)\big)$ are contained within the unit circle. This implies
\begin{align}
\mathbf{E}\Big(L(k+1)\big|x(k)\Big)&\leq (x(k)-x_{\rm ave}\mathbf{1}\big)^T \mathbf{E}\big(\Psi^2(k)\big) \big(x(k)-x_{\rm ave}\mathbf{1}\big)\nonumber\\
&\leq \Big(1-2\Big( T_k(1-T_k)\alpha-S_k(1+S_k)\gamma \Big)\frac{1}{n} \lambda_2^\ast \Big)L(k)\nonumber\\
&=\Big(1- \frac{2}{n}  \mathcal{I}_k \Big)L(k).\nonumber
\end{align}

\item[(ii).] Suppose $T_k(1-T_k)< S_k(1+S_k)$. Then we have
\begin{align}
1\leq \lambda_i\Big( \mathbf{E}\big(\Psi^2(k)\big)\Big)\leq  1-2\Big( T_k(1-T_k)\alpha-S_k(1+S_k)\gamma \Big)\frac{1}{n} \lambda_n^\ast\nonumber
\end{align}
for all eigenvalues of $\mathbf{E}\big(\Psi^2(k)\big)$, which also yields
\begin{align}
\mathbf{E}\Big(L(k+1)\big|x(k)\Big)&\leq \Big(1- \frac{2}{n}  \mathcal{I}_k \Big)L(k).\nonumber
\end{align}
\end{itemize}
Therefore, repeating the analysis in the proof of Proposition  \ref{thm2}, we obtain
\begin{align}
\mathbf{E}\Big(\limsup_{k\rightarrow\infty}L(k)\Big)\leq\limsup_{k\rightarrow\infty} \mathbf{E}\Big(L(k)\Big)=0,\nonumber
\end{align}
as long as $\prod_{k=0}^\infty \Big(1- \frac{2}{n}\mathcal{I}_k\Big)=0$.   This immediately implies
\begin{align}
\mathbf{P}\Big(\lim_{k\rightarrow\infty} x_i(k)=x_{\rm ave}\Big)=1,\nonumber
\end{align}
which completes the proof. \hfill$\square$

\vspace{2mm}

Next, we discuss the state disagreement under symmetric updates. The following conclusion holds on the state disagreement in expectation.
\begin{prop}\label{thm5}
Suppose {\bf A1}  and {\bf A5} hold.   Disagreement convergence is achieved in expectation, i.e., $\lim_{k\rightarrow \infty}\mathbf{E}\big( \mathcal{H}(k)\big)=\infty$,  for almost all initial values if
\begin{align}
\prod_{k=0}^\infty \Big(1- \frac{2}{n}\hat{\mathcal{I}}_k\Big)=\infty,\nonumber
\end{align}
where
\begin{equation}
	\hat{\mathcal{I}}_k =
	\begin{cases}
		\Big( T_k(1-T_k)\alpha-S_k(1+S_k)\gamma\Big)\lambda_n^\ast, & \text{if $T_k(1-T_k)\geq S_k(1+S_k)$;}\\
\Big( T_k(1-T_k)\alpha-S_k(1+S_k)\gamma\Big) \lambda_2^\ast, & \text{if $T_k(1-T_k)< S_k(1+S_k)$.}\\
	\end{cases}
\end{equation}
\end{prop}
\noindent {\it Proof.} By establishing the upper bound of the right-hand side of Eq. (\ref{25}), we obtain
\begin{align}
\mathbf{E}\Big(L(k+1)\big|x(k)\Big)&\geq \Big(1- \frac{2}{n}  \hat{\mathcal{I}}_k \Big)L(k) \nonumber
\end{align}
for all $k$, which implies
\begin{align}\label{27}
\mathbf{E}\Big(L(k+1)\Big)\geq  \Big(1- \frac{2}{n}\hat{\mathcal{I}}_k\Big)\mathbf{E}\Big(L(k)\Big).
\end{align}
This implies the desired conclusion straightforwardly. \hfill$\square$

\vspace{2mm}

For a.s.   disagreement divergence, we present the following result.
\begin{prop}\label{thm6}
Suppose {\bf A1}  and {\bf A5} hold.   Disagreement divergence is achieved a.s. for almost all initial conditions if

(i)   there exists a constant $S^\ast>0$ such that $S_k\leq S^\ast$ for all $k$;

(ii)   there exists a constant   $0<\varepsilon<1/2$ such that either $T_k\in[0,1/2-\varepsilon]$ or $T_k\in[1/2+\varepsilon,1]$  for all $k$;

(iii) there exists $0<\tau<1$ such that $\limsup_{m\rightarrow \infty} \sum_{k=0}^m \mathcal{J}_\tau(k)=O(m) $, where
$$
\mathcal{J}_\tau(k)=\log \Big[ \big(1+4\tau  (S_k^2+S_k)\big)^{p_k}\big(2T_k-1\big)^{2\alpha}\Big]
$$
with $p_k=-\frac{\frac{2}{n}\hat{\mathcal{I}}_k+\gamma\big(1+4\tau (S_k^2+S_k)\big) }{4(1-\tau)(S_k^2+S_k)}$, and by definition $b_k=O(c_k)$ means that $\limsup_{k\rightarrow\infty}{b(k)}/{c(k)}<\infty$ is a nonzero  constant.
\end{prop}
{\it Proof.} We divide the proof into three steps.

\noindent {Step 1.} In this step, we show that with probability one and for almost all initial conditions, finite-time agreement convergence  cannot be achieved.  According to  (\ref{s3}), we obtain
\begin{align}
\mathbf{P}\Big( \mathcal{H}(k+1)\geq \big(1- 2T_k\big) \mathcal{H}(k)\Big)=1 \nonumber
\end{align}
for all $k\geq0$ if $T_k\in[0,1/2-\varepsilon]$. Observing that $1- 2T_k\geq 2\varepsilon>0$
we see that $\mathcal{H}(k)>0$ for all $k$ with probability one for all initial values satisfying $\mathcal{H}(k_0)>0$. This  holds also for the other case $T_k\in[1/2+\varepsilon,1]$ based on a symmetric argument.

Suppose nodes $u$ and $v$ reach the maximum and minimum values at time $k$, respectively, i.e.,
$$x_u(k)=\max_{i\in\mathcal{V}}x_{i}(k); \quad x_v(k)=\min_{i\in \mathcal{V}}x_{i}(k).
$$
Then we have
\begin{align}
L(k)=\sum_{i=1}^n |x_i(k)-x_{\rm ave}|^2&\geq |x_u(k)-x_{\rm ave}|^2+|x_v(k)-x_{\rm ave}|^2\geq \frac{1}{2}|x_u(k)-x_v(k)|^2=\frac{1}{2}\mathcal{H}^2(k),\nonumber
\end{align}
which implies $L(k)>0$ with probability one for almost all initial conditions.

Therefore, with probability one, we can introduce a sequence of random variables $\{\varpi_k\}_0^\infty$ satisfying
$$
L(k+1)=\varpi_k L(k),\ k\geq0,
$$
and we see from (\ref{27}) that
\begin{align}\label{36}
 \mathbf{E}\big( \varpi_k \big)=\mathbf{E}\big(L(k+1)\big)/ \mathbf{E}\big(L(k)\big)
 \geq  1- \frac{2}{n}\hat{\mathcal{I}}_k
 \doteq Z_k.
\end{align}

\vspace{3mm}

\noindent {Step 2.} We establish a lower bound for $\mathbf{E}\big(\log \varpi_k \big)$ in this step.

Recall that $\Psi(k)$ is the random matrix introduced in (\ref{34}). It is not hard to find that for every possible sample, $\Psi_{\langle ij\rangle}^+$ or $\Psi_{\langle ij\rangle}^-$ of $\Psi(k)$, it holds that
\begin{align}\label{35}
\min\Big\{ |\lambda_i|:\ \lambda_i\in \sigma(\Psi_{\langle ij\rangle}^+)\cup\sigma(\Psi_{\langle ij\rangle}^-)\Big\}\geq \min\Big\{ |\lambda_i|:\ \lambda_i\in \sigma( V_k)\Big\} =2T_k-1,
\end{align}
where
\begin{equation}
V_k=\left(
\begin{array}{ccc}
 1-T_k & T_k  \\
 T_k & 1-T_k
\end{array}
\right).
\end{equation}
Noticing that
\begin{align}
L(k+1)= (x(k)-x_{\rm ave}\mathbf{1}\big)^T\Psi^2(k) \big(x(k)-x_{\rm ave}\mathbf{1}\big)\geq \min_{\lambda_i\in \sigma(\Psi(k))} |\lambda_i|^2 L(k), \nonumber
\end{align}
the definition of $\varpi_k$ and (\ref{35}) yield
\begin{align}\label{38}
\mathbf{P}\Big(\varpi_k\geq (2T_k-1)^2\Big)=\mathbf{P}\Big(\log \varpi_k\geq \log(2T_k-1)^2\Big)=1.
\end{align}

Similarly, observing that
\begin{align}
\max\Big\{ |\lambda_i|:\ \lambda_i\in \sigma(\Psi_{\langle ij\rangle}^+)\cup\sigma(\Psi_{\langle ij\rangle}^-)\Big\}\leq \max\Big\{ |\lambda_i|:\ \lambda_i\in \sigma( \hat{V}_k)\Big\} =2S_k+1,
\end{align}
where
\begin{equation}
\hat{V}_k=\left(
\begin{array}{ccc}
 1+S_k & -S_k  \\
 -S_k & 1+S_k
\end{array}
\right), \nonumber
\end{equation}
we obtain
\begin{align}\label{39}
\mathbf{P}\Big(\varpi_k\leq (2S_k+1)^2\Big)=\mathbf{P}\Big(\log \varpi_k\leq \log(2S_k+1)^2\Big)=1.
\end{align}

\vspace{2mm}

Noticing (\ref{36}) and that
\begin{align}
 \mathbf{E}\big( \varpi_k \big)=\int_{\varpi_k\leq 1}\varpi_k+\int_{\varpi_k> 1}\varpi_k\leq 1+\int_{\varpi_k> 1}\varpi_k, \nonumber
\end{align}
we obtain
\begin{align}
\int_{\varpi_k> 1}\varpi_k\geq  \mathbf{E}\big( \varpi_k \big)-1\geq Z_k-1. \nonumber
\end{align}

\vspace{2mm}

Take $0<\tau<1$  a  constant.   Based on the definition of the desired algorithm, we see that
\begin{align}
 \mathbf{P}\Big(\varpi_k> 1\Big)\leq \mathbf{P}\Big( \mathscr{R}_{ij}(k)\ \mbox{happens for some node pair $(i,j)$}\Big)=\gamma. \nonumber
\end{align}
Now we conclude that
\begin{align}\label{37}
Z_k-1\leq\int_{\varpi_k> 1}\varpi_k\leq \hat{p}_k(2S_k+1)^2+ \big(1-\tau+\tau  (2S_k+1)^2\big)(\gamma-\hat{p}_k),
\end{align}
where by definition
$$
\hat{p}_k\doteq\mathbf{P}\big(1-\tau+\tau  (2S_k+1)^2\leq\varpi_k\leq  (2S_k+1)^2\big).
$$
After some simple algebra we see from (\ref{37}) that
\begin{align}\label{40}
\hat{p}_k \geq \frac{Z_k-1-\gamma\big(1-\tau+\tau(2S_k+1)^2 \big)}{4(1-\tau)(S_k^2+S_k)}=-\frac{\frac{2}{n}\hat{\mathcal{I}}_k+\gamma\big(1+4\tau (S_k^2+S_k)\big) }{4(1-\tau)(S_k^2+S_k)}\doteq  p_k.
\end{align}

Combining (\ref{38}), (\ref{39}) and (\ref{40}), we  eventually arrive at the following lower bound of  $\mathbf{E} \log \varpi_k$:
\begin{align}\label{42}
\mathbf{E} \log \varpi_k &\geq \hat{p}_k\log\Big(1-\tau+\tau  (2S_k+1)^2\Big)+\alpha\log(2T_k-1)^2\nonumber\\
&\geq \log \Big[\big(1+4\tau  (S_k^2+S_k)\big)^{p_k}\big(2T_k-1\big)^{2\alpha}\Big]\nonumber\\
&\doteq  \mathcal{J}_\tau(k).
\end{align}

\vspace{3mm}

\noindent{Step 3.} In this step, we complete the final piece of the proof by a contradiction argument. Suppose there exist two constants $M_0\geq 0$ and $0<p<1$ such that
\begin{equation}
\mathbf{P}\Big(\limsup_{k\rightarrow \infty} \mathcal{H}(k)\leq M_0\Big)=p.
\end{equation}
Noticing that
\begin{align}
L(k)=\sum_{i=1}^n|x_i(k)-x_{\rm ave}|^2\leq n\mathcal{H}^2(k),\nonumber
\end{align}
we further conclude
\begin{equation}
\mathbf{P}\Big(\limsup_{k\rightarrow \infty} L(k)\leq n M^2_0\Big)\geq p, \nonumber
\end{equation}
which yields
\begin{equation}
\mathbf{P}\Big(\limsup_{m\rightarrow \infty} \log L(m+1)=\sum_{k=0}^m\log\varpi_k \leq \log \big(n M^2_0\big)\Big)\geq p. \nonumber
\end{equation}
This leads to
\begin{align}\label{41}
\mathbf{P}\Big( \lim_{m\rightarrow \infty} \frac{\sum_{k=0}^m\log\varpi_k}{m}\leq 0\Big)\geq p.
\end{align}

On the other hand, noting that the node updates are independent of time and node states, and that $\mathbf{V} (\log \varpi_k)$ is bounded according to (\ref{38}) and (\ref{39}),  the Strong Law of Large Numbers  and (\ref{42}) suggest that
\begin{align}
\mathbf{P}\Big( \lim_{m\rightarrow \infty} \frac{1}{m}\sum_{k=0}^m\big(\log\varpi_k- \mathcal{J}_\tau(k)\big) \geq 0\Big)\geq \mathbf{P}\Big( \lim_{m\rightarrow \infty} \frac{1}{m}\sum_{k=0}^m\big(\log\varpi_k-\mathbf{E}\log \varpi_k\big) = 0\Big)=1,\nonumber
\end{align}
which contradicts (\ref{41}) if $\limsup_{m\rightarrow \infty} \sum_{k=0}^m \mathcal{J}_\tau(k)=O(m) $.

The desired conclusion thus follows and this completes the proof. \hfill$\square$

\vspace{3mm}

We conclude this subsection by the following conclusion under the condition when $T_k$ and $S_k$ are time-invariant, which follows straightforwardly from Propositions \ref{thm4}, \ref{thm5} and \ref{thm6}.

\begin{thm}\label{beer}
Suppose {\bf A1}  and {\bf A5} hold. Let $T_\star\in[0,1]$ and $ S_\star >0$ be two given constants. Assume that $T_k\equiv T_\star$ and $S_k\equiv S_\star$. Then
 $$
{D}_0=S_\star(1+S_\star)\gamma-T_\star(1-T_\star)\alpha
 $$
is a critical convergence measure regarding  the state convergence of the considered network. To be precise, we have
\begin{itemize}
\item[(i)] Global agreement convergence is achieved a.s. if ${D}_0<0$;

\item[(ii)] Disagreement divergence  is achieved in expectation for almost all initial values  if ${D}_0>0$;

\item[(iii)] State oscillation  is achieved in expectation, i.e., $\mathbf{E}\big(L(k)\big)=L(k_0)$ for all $k\geq k_0$   if ${D}_0=0$;

    \item[(iv)] Disagreement divergence is achieved a.s. for almost all initial conditions if $T_\star\neq 1/2$ and ${D}_0$ is sufficiently large, i.e.,  there exists $0<\tau<1$ such that
$$
\big(1+4\tau  (S_\star^2+S_\star)\big)^{p^\ast}\big(2T_\star-1\big)^{2\alpha}>1,
$$
where
$$
p^\ast=\frac{2{D}_0 \lambda_2^\ast-n\gamma\big(1+4\tau (S_\star^2+S_\star)\big) }{4n(1-\tau)(S_\star^2+S_\star)}.
$$
\end{itemize}
\end{thm}

\begin{remark}
It is surprising that the convergence measure  ${D}_0$ in  Theorem \ref{beer} does not rely on the network topology. This is to say, if all the nodes may misbehave with equal  probability as the proposed algorithm, then there is no particular topology which can be viewed as ``better" than others  in terms of agreement convergence.
\end{remark}

\subsection{Asymmetric Update}
In this subsection, we discuss  asymmetric node updates. We introduce the following assumption.

\vspace{2mm}
\noindent {\bf A6.} {\em (Asymmetric Update)} Both $\mathscr{A}_{ij}(k)\bigcap \mathscr{A}_{ji}(k)$ and $\mathscr{R}_{ij}(k)\bigcap\mathscr{R}_{ji}(k)$  are trivial events for all $(i,j)$ and $k$.

\vspace{2mm}

The main result on a.s. agreement convergence  under asymmetric update is  as follows.
\begin{prop}\label{thm7}
Suppose {\bf A1}  and {\bf A6} hold.  Global  agreement convergence  is achieved  a.s.  if  there exists  $S^\ast>0$ a such that $S_k\leq S^\ast$ for all $k$ and
\begin{align}\label{46}
&\prod_{k=0}^\infty \Bigg[ 1-\Big(\frac{\alpha a_\ast }{n}\Big)^{n-1}  \hat{T}_k+\Big(1-\big(1-\gamma\big)^{n-1}\Big)\Big(\hat{S}_k-1\Big)\Bigg]=0,
\end{align}
where
$$
 \hat{T}_k=\prod_{m=k(n-1)}^{(k+1)(n-1)-1}T_m(1-T_{m}); \quad \hat{S}_k=\prod_{m=k(n-1)}^{(k+1)(n-1)-1}\big(S_{m}+1\big).
$$
\end{prop}
{\it Proof.} Following from the proof of Proposition \ref{thm3}, we have
\begin{align}\label{43}
\mathbf{P}\Big(\mathcal{H}(k_\ast+n-1) \leq \Big(1-\prod_{k=k_\ast}^{k_\ast+n-2}T_{k}(1-T_{k})\Big)\mathcal{H}(k_\ast)\Big)\geq \Big(\frac{\alpha a_\ast }{n}\Big)^{n-1}
\end{align}
for all $k_\ast\geq0$.

On the other hand, the definition of the randomized algorithm leads to
\begin{align}\label{44}
\mathbf{P}\Bigg(\mathcal{H}(k_\ast+n-1) \leq \Big(\prod_{k=k_\ast}^{k_\ast+n-2}\big(S_{k}+1\big)\Big)\mathcal{H}(k_\ast)\Bigg)=1
\end{align}
and
\begin{align}\label{45}
\mathbf{P}\Big(\mathcal{H}(k_\ast+n-1) > \mathcal{H}(k_\ast)\Big)\leq1-\big(1-\gamma\big)^{n-1}
\end{align}
since $\mathcal{H}(k_\ast+n-1) > \mathcal{H}(k_\ast)$ implies  that repulsion happens at least one time during $[k_\ast,k_\ast+n-1)$.

We conclude from (\ref{43}) and (\ref{44})  that
\begin{align}
\mathbf{E}\big(\mathcal{H}(k_\ast+n-1)\big) &\leq \Bigg[ \Big(1- \prod_{k=k_\ast}^{k_\ast+n-2}T_k(1-T_{k})\Big)\Big(\frac{\alpha a_\ast }{n}\Big)^{n-1}+\big(1-\gamma\big)^{n-1}-\Big(\frac{\alpha a_\ast }{n}\Big)^{n-1}\nonumber\\
&\quad \quad \quad \quad \quad \quad \quad \quad \quad \quad +\Big(1-\big(1-\gamma\big)^{n-1}\Big)\prod_{k=k_\ast}^{k_\ast+n-2}\big(S_{k}+1\big)\Bigg] \mathbf{E}\big(\mathcal{H}(k_\ast)\big)\nonumber\\
&=\Bigg[ 1- \Big(\frac{\alpha a_\ast }{n}\Big)^{n-1}\prod_{k=k_\ast}^{k_\ast+n-2}T_k(1-T_{k})\nonumber\\
&\quad \quad \quad \quad \quad \quad \quad \quad \quad \quad +\Big(1-\big(1-\gamma\big)^{n-1}\Big)\Big(\prod_{k=k_\ast}^{k_\ast+n-2}\big(S_{k}+1\big)-1\Big)\Bigg] \mathbf{E}\big(\mathcal{H}(k_\ast)\big)\nonumber
\end{align}
for all $k_\ast>0$. This implies
\begin{align}
\mathbf{E}\Big(\lim_{s\rightarrow \infty}\mathcal{H}\big(s(n-1)\big)\Big)\leq \lim_{s\rightarrow \infty}\mathbf{E}\Big(\mathcal{H}\big(s(n-1)\big)\Big)=0
\end{align}
if (\ref{46}) holds, and thus
\begin{align}\label{47}
\mathbf{P}\Big(\lim_{k\rightarrow \infty}\mathcal{H}\big(k(n-1)\big)=0\Big)=1.
\end{align}
Since  there exists  $S^\ast>0$ a such that $S_k\leq S^\ast$ for all $k$, we see from (\ref{44}) and (\ref{47}) that
\begin{align}
\mathbf{P}\Big(\lim_{k\rightarrow \infty}\mathcal{H}\big(k\big)=0\Big)=1. \nonumber
\end{align}
The desired conclusion follows. \hfill$\square$

\vspace{3mm}
Next, we study a.s.  disagreement divergence. The following conclusion holds.
\begin{prop}\label{thm8}
Suppose {\bf A1}  and {\bf A6} hold.  Disagreement divergence is achieved  a.s. for almost all initial values  if

(i)   there exist two constants $S^\ast>0$ and  $0<T^\ast<1$ such that $S_k\leq S^\ast$ and $T_k\leq T^\ast$ for all $k$.

(ii) there exists an integer $Z\geq0$ such that $\sum_{k=0}^m \mathcal{J}_Z(k)=O(m)$, where
$$
\mathcal{J}_Z(k)=\Big(\frac{\gamma a_\ast}{n}\Big)^{Z+1}\log \Big(\frac{1}{n-1} \prod_{\varsigma=k(Z+1)}^{(k+1)(Z+1)-1}\big(1+S_\varsigma\big)\Big)+\big(1-(1-\alpha)^{Z+1}\big) \log \prod_{\varsigma=k(Z+1)}^{(k+1)(Z+1)-1}\big(1-T_\varsigma\big).
$$
\end{prop}
{\it Proof.}  Suppose node pair $(i,j)$ is selected at time $k$. According to  the definition of the considered randomized algorithm, we obtain
\begin{equation}\label{60}
	\big|x_i(k+1)-x_j(k+1)\big|=
	\begin{cases}
		\big|x_i(k)-x_j(k)\big|, & \text{if $\mathscr{N}_{ij}(k)$ happens;}\\
(1-T_k)\big|x_i(k)-x_j(k)\big|, & \text{if $\mathscr{A}_{ij}(k)$ happens;}\\
(1+S_k)\big|x_i(k)-x_j(k)\big|, & \text{if $\mathscr{R}_{ij}(k)$ happens.}
	\end{cases}
\end{equation}
Therefore, with assumption {\bf A6}, we obtain
\begin{align}
\mathbf{P}\Big( \mathcal{H}(k+1)\geq \big(1- T_k\big) \mathcal{H}(k)\Big)\geq \mathbf{P}\Big( \mathcal{H}(k+1)\geq \big(1- T^\ast\big) \mathcal{H}(k)\Big)=1 \nonumber
\end{align}
for all $k\geq0$. This implies for all initial values satisfying $\mathcal{H}(k_0)>0$, agreement convergence is achieved only in infinite time with probability one. As a result, we can well define a sequence of random variable, $\{\hat{\varpi}_k\}_0^\infty$, such that
$$
\mathcal{H}(k+1)=\hat{\varpi}_k \mathcal{H}(k),\ k\geq0.
$$

Now with (\ref{60}), it is straightforward to  conclude that
\begin{align}\label{61}
\mathbf{P} \Big(\hat{\varpi}_k\geq 1- T_k\Big)=1
\end{align}
and
\begin{align}\label{65}
\mathbf{P} \Big(\hat{\varpi}_k< 1\Big)\leq \alpha.
\end{align}
Moreover, based on the weak connectivity assumption {\bf A1}, for any $k\geq 0$, there always exist two nodes $i_0$ and $j_0$ such that either $a_{i_0j_0}>0$ or $a_{j_0i_0}>0$, and $$
\big|x_{i_0}(k)-x_{j_0}(k)\big|\geq \frac{1}{n-1}\mathcal{H}(k).
$$
Note that if  $a_{i_0j_0}>0$ or $a_{j_0i_0}>0$, and $
\big|x_{i_0}(k)-x_{j_0}(k)\big|\geq \mu\mathcal{H}(k)$ for some $\mu>0$, we have $\big|x_{i_0}(k+1)-x_{j_0}(k+1)\big|\geq (1+S_k)\mu \mathcal{H}(k)$ with probability ${\gamma a_\ast}/{n}$.

Thus, the case with $\mathscr{R}_{ij}(k)$ happening in (\ref{60}) leads to
\begin{align}
\mathbf{P} \Big(\hat{\varpi}_k\geq \frac{1+S_k}{n-1}\Big)\geq \frac{\gamma a_\ast}{n},
\end{align}
and
\begin{align}\label{62}
\mathbf{P} \Big(\hat{\varpi}_{k+s}\cdots \hat{\varpi}_k\geq \frac{1}{n-1} \prod_{\varsigma=k}^{k+s}\big(1+S_\varsigma\big)\Big)\geq  \Big(\frac{\gamma a_\ast}{n}\Big)^{s+1}, \ \ s\geq 0
\end{align}
recalling that $a_\ast=\min\{a_{ij}:\ a_{ij}>0\}$ is the lower bound of the nonzero entries of $A$.

Therefore, letting $Z\geq0$ be an integer, we can eventually conclude  from (\ref{61}), (\ref{65}) and (\ref{62}) that
\begin{align}
\sum_{k=k_\ast}^{k_\ast+Z}\mathbf{E}\log\hat{\varpi}_k &\geq  \Big(\frac{\gamma a_\ast}{n}\Big)^{Z+1}\log \Big(\frac{1}{n-1} \prod_{k=k_\ast}^{k_\ast+Z}\big(1+S_k\big)\Big)+\big(1-(1-\alpha)^{Z+1}\big) \log \prod_{k=k_\ast}^{k_\ast+Z}\big(1-T_k\big).\nonumber
\end{align}
The desired conclusion follows from the same argument as the proof of Proposition  \ref{thm6} based on the Strong Law of Large Numbers. This completes the proof. \hfill$\square$

\vspace{3mm}

We also end the discussion of  this subsection by a theorem for the case when $T_k$ and $S_k$ are time-invariant.  Applying the same analysis  methods of proving  Propositions \ref{thm7} and \ref{thm8}, we obtain the following result.

\begin{thm}\label{prop4}
Suppose {\bf A1}  and {\bf A6} hold. Let $T_\star\in[0,1]$ and $ S_\star >0$ be two given constants. Assume that $T_k\equiv T_\star$ and $S_k\equiv S_\star$. Then we have
\begin{itemize}
\item[(i)] Global agreement convergence is achieved a.s. if \begin{align}
  \Big(1-\big(1-\gamma\big)^{n-1}\Big)\Big(\big(S_{\ast}+1\big)^{n-1}-1\Big)<\Big(\frac{\alpha a_\ast }{n}\Big)^{n-1}  \Big(\max \big\{{T}_\star,1-T_\star\big\}\Big)^{n-1};\nonumber
\end{align}

    \item[(ii)] Disagreement divergence is achieved a.s. for almost all initial conditions if  there exists  an integer $Z\geq 0$ such that
$$
\Big(\frac{\gamma a_\ast}{n}\Big)^{Z+1}\log \frac{S_\star^{Z+1}}{n-1}+{\big(1-(1-\alpha)^{Z+1}\big)}\big(Z+1\big) \log \big(1-T_\star\big)>0.
$$
\end{itemize}
\end{thm}

\begin{remark}
It is unclear from Theorems \ref{beer} and \ref{prop4} if  symmetric  or  asymmetric updates are better guiding  the network states to agreement or  disagreement. In order to answer this question,  more accurate estimates of the state evolution are needed. We guess  the answer will highly depend on the network topology.
\end{remark}

\subsection{Numerical Example}
We present a numerical example in order to illustrate the critical measure established in Theorem \ref{beer}.

\begin{figure}
\centerline{\epsfig{figure=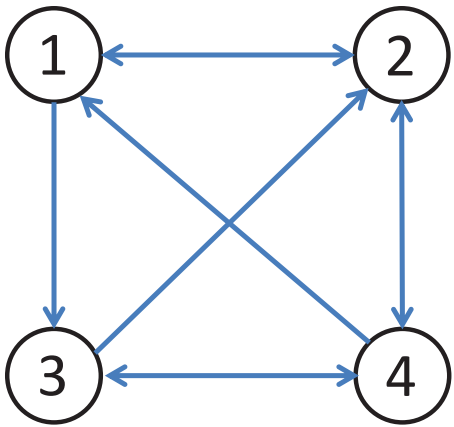, width=0.40\linewidth=0.25}}
\caption{The underlying communication graph. }\label{graph}
\end{figure}

Consider four nodes $1,\dots,4$. The node meeting probability matrix is given by
\begin{align}
A=[a_{ij}]=\begin{pmatrix}
0 & 1/2 & 0 & 1/2 \\
1/2 & 0 & 1/4 & 1/4 \\
1/3 & 0 & 0 & 2/3 \\
0 & 1/3 & 2/3 & 0
\end{pmatrix}.\nonumber
\end{align}
The underlying graph $\mathcal{G}_0$ is shown in the Fig. \ref{graph}. The initial values are taken as $x_i(0)=i, i=1,\dots,4$. We take $\alpha=\beta=\gamma=1/3$ and let $T_k\equiv T_\star$ and $S_k\equiv S_\star$.

\begin{figure}
\centerline{\epsfig{figure=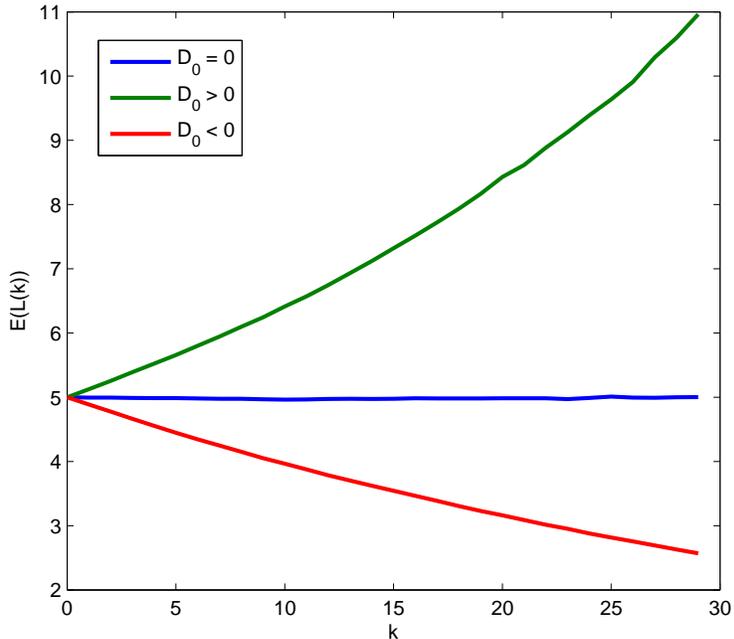, width=0.70\linewidth=0.25}}
\caption{The expected value of $L(k)$ for different $D_0$. }\label{numerical}
\end{figure}

Take $T_\star=1/4$ and $S_\star=(\sqrt{7}-2)/4,\ (\sqrt{7}-2)/4-0.05,\ (\sqrt{7}-2)/4+0.05$, respectively.  The corresponding values of $D_0=S_\star(1+S_\star)\gamma-T_\star(1-T_\star)\alpha$ are then given by $0$, $-0.0212$, and $0.0229$. We run the considered randomized algorithm for $10^5$ times, and then take the average value of the consensus measure $L(k)=\sum_{i=1}^4\big(x_i(k)-x_{\rm ave}\big)^2$  as the empirical  estimate of the expected value of $L(k)$. The transition of $\mathbf{E}(L(k))$ for these three cases of  $D_0$ is shown in Fig. \ref{numerical}. The numerical result is consistent with the conclusion in Theorem \ref{beer}.

\section{Conclusions}
This paper proposed a model for investigating  node misbehavior   in  distributed information processing over random networks.  At each instance, two nodes were selected for a meeting with a given probability. When nodes meet, there were three events for the node update: attraction, neglect, or repulsion. Attraction event follows the standard averaging algorithm targeting a consensus; neglect event means the selected node will stick to its current state; repulsion event represents the case when nodes are against the consensus convergence. Each node was assumed to follow one of these three update rules at random. Both symmetric and asymmetric node updates were studied. After obtaining two general impossibility theorems, a series of necessary  and/or  sufficient  conditions were established for the network to reach a.s. agreement convergence, or  a.s. disagreement divergence.  For the symmetric update model, we established a critical convergence  measure stating that convergence transits from agreement to disagreement whenever this measure goes from negative to positive. The proposed algorithm may serve  a uniform model for characterizing node misbehavior in  communication network, large-scale control system, or social networks. To the best of our knowledge,  the obtained results for the first time in the literature gave a clear description on the possible disagreement divergence for distributed averaging due to node misbehavior. More challenges lie in the optimal policy  for the nodes to take bad action from a tradeoff between the risk of being discovered and the result it generates, and the case when bad action only takes place for some particular neighboring relations.


\begin{thebibliography}{99}


\bibitem{lat} G. Latouche, V. Ramaswami. {\em  Introduction to Matrix Analytic Methods in Stochastic Modeling.} 1st edition, ASA SIAM, 1999.


\bibitem{god}
C. Godsil and G. Royle.
\newblock {\em Algebraic Graph Theory.}
\newblock New York: Springer-Verlag, 2001.

\bibitem{er}P. Erd\"{o}s and A. R\'{e}nyi, ``On the evolution of random graphs," {\em Publications of the Mathematical Institute of the Hungarian Academy of Sciences}, pp. 17-61, 1960.

\bibitem{watts}  D.J. Watts and S.H. Strogatz, ``Collective dynamics of small-world networks," {\em Nature} vol. 393 no. 6684, pp. 409-410, 1998.

\bibitem{bara} A.-L. Barab\'{a}si and R. Albert, ``Emergence of scaling in random networks," {\em Science}, vol. 286, no. 5439, pp. 509-512, 1999.

\bibitem{rg1} S. N. Dorogovtsev,  and  J. F. F. Mendes. {\em  Evolution of Networks: From Biological Nets to the Internet and WWW}. Oxford Univ. Press, 2003.

\bibitem{rg2} M. Newman, A.-L. Barab\'{a}si  and D. J. Watts. {\em  The Structure and Dynamics of Networks.}  Princeton Univ. Press, 2006.

\bibitem{rg3}A. Barrat, M. Barth\'{e}lemy, and A. Vespignani. {\em Dynamical processes in complex networks.} Cambridge University Press, 2008.

\bibitem{boid} C.W.  Reynolds, ``Flocks, herds and schools: A distributed behavioral model," {\em Computer Graphics}, 21(4), pp. 25-34, 1987.


\bibitem{vic95}
T. Vicsek, A. Czirok, E. B. Jacob, I. Cohen, and O. Schochet.
\newblock Novel type of phase transitions in a system of self-driven
particles.
\newblock{\em Physical Review Letters}, vol. 75, 1226-1229, 1995.



\bibitem{control1} A. Rahmani, M. Ji, M. Mesbahi, and M. Egerstedt. ``Controllability of multi-agent systems: from a graph-theoretic perspective," {\em SIAM Journal on Control and Optimization}, vol. 48, no. 1, pp. 162-186,  2009.

    \bibitem{control2} Y.-Y. Liu, J.-J. Slotine and  A.-L. Barab\'{a}si, ``Controllability of complex networks," {\em Nature}, 473, pp. 167-173, 2011.

\bibitem{cs2} S. Muthukrishnan, B. Ghosh, and M. Schultz, ``First and second order
diffusive methods for rapid, coarse, distributed load balancing,"  {\it Theory
of Computing Systems}, vol. 31, pp. 331-354, 1998.

\bibitem{cs3} R. Diekmann, A. Frommer, and B. Monien, ``Efficient schemes for
nearest neighbor load balancing," {\it Parallel Computing}, vol. 25, pp. 789-812, 1999.



\bibitem{saber04}
R. Olfati-Saber and R. Murray, ``Consensus problems in the networks of agents with switching topology
and time dealys,"
\newblock {\em IEEE Trans. Autom.
Control}, vol. 49, no. 9, pp. 1520-1533, 2004.

\bibitem{mar}
S. Martinez, J. Cort\'{e}s, and F. Bullo, ``Motion coordination
with distributed information,"
\newblock {\em IEEE Control Systems Magazine}, vol. 27, no. 4, pp. 75-88, 2007.




\bibitem{moura1}  A. G. Dimakis,
S. Kar,  J. M. F. Moura,  M. G. Rabbat,  and A. Scaglione, ``Gossip algorithms for distributed signal processing,"  {\em Proceedings of IEEE}, vol. 98, no. 11, pp. 1847-1864, 2010.

\bibitem{moura2}S. Kar and J. M. F. Moura, ``Convergence rate analysis of distributed gossip
(linear parameter) estimation: fundamental
limits and tradeoffs," {\em IEEE Journal of Selected Topics in Signal Processing}, vol. 5, no. 4, pp.674-690, 2011.


\bibitem{social1} M. H. DeGroot, ``Reaching a consensus," {\it Journal of the American Statistical Association}, vol. 69, no. 345, pp. 118-121, 1974.

\bibitem{social2} P. M. DeMarzo, D. Vayanos, J. Zwiebel, ``Persuasion bias, social influence, and unidimensional
opinions," {\em Quarterly Journal of Economics}, vol. 118, no. 3, pp. 909-968, 2003.

\bibitem{social3} B. Golub and M. O. Jackson, ``Na\"{i}ve learning in social networks and the wisdom of crowds," {\it American Economic Journal: Microeconomics}, vol. 2, no. 1, pp. 112-149, 2007.



\bibitem{haj} J. Hajnal, ``Weak ergodicity in non-homogeneous Markov chains," {\em Proc. Cambridge
Philos. Soc.}, no. 54, pp. 233-246, 1958.


\bibitem{wolf}J. Wolfowitz, ``Products of indecomposable, aperiodic, stochastic matrices,"
{\em Proc. Amer. Math. Soc.}, vol. 15, pp. 733-736, 1963.



\bibitem{tsi}
J. Tsitsiklis, D. Bertsekas, and M. Athans, ``Distributed asynchronous
deterministic and stochastic gradient optimization algorithms," {\em
IEEE Trans. Autom. Control}, vol. 31, pp. 803-812, 1986.




\bibitem{jad03}
A. Jadbabaie, J. Lin, and A. S. Morse,
``Coordination of groups of mobile autonomous agents using nearest neighbor rules,"  {\em IEEE Trans. Autom.Control}, vol. 48, no. 6, pp. 988-1001, 2003.


\bibitem{tsi2} A. Nedi\'{c}, A. Olshevsky, A. Ozdaglar, and J. N. Tsitsiklis, ``On distributed
averaging algorithms and qantization effects," {\it IEEE Trans.
Autom. Control}, vol. 54, no. 11, pp.  2506-2517, 2009.


\bibitem{caoming1} M. Cao,  A. S. Morse and B. D. O. Anderson, ``Reaching a consensus in a dynamically changing
environment: a graphical approach," \newblock {\em SIAM J. Control Optim.}, vol. 47, no. 2, 575-600, 2008.

\bibitem{caoming3}  M. Cao,  A. S. Morse and B. D. O. Anderson, ``Agreeing asynchronously," {\em IEEE Trans. Autom. Control}, vol. 53, no. 8, 1826-1838, 2008.



\bibitem{mor}
L. Moreau, ``Stability of multi-agent systems with time-dependent
communication links," {\em IEEE Trans. Autom. Control}, vol. 50,
pp. 169-182, 2005.

\bibitem{ren} W. Ren and R. Beard, ``Consensus seeking in multi-agent systems under dynamically changing interaction topologies," {\em IEEE Trans. Autom. Control}, vol. 50, no. 5, pp. 655-661, 2005.

\bibitem{julien1} V. D. Blondel, J. M. Hendrickx and J. N. Tsitsiklis, ``Continuous-time average-preserving opinion dynamics with opinion-dependent communications," {\em SIAM Journal on Control and Optimization}, vol. 48, no. 8, pp. 5214-5240, 2010.

\bibitem{julien2} V. D. Blondel, J. M. Hendrickx and J. N. Tsitsiklis, ``On Krause's multi-agent consensus model with state-dependent connectivity," {\em IEEE Transactions on Automatic Control}, vol. 54, no. 11, pp. 2586-2597, 2009.



\bibitem{hatano} Y. Hatano and M. Mesbahi, ``Agreement over random networks,"
{\em IEEE Trans. on Autom. Control}, vol. 50, no. 11, pp. 1867-1872,
2005.

\bibitem{wu} C. W. Wu, ``Synchronization and convergence of linear dynamics in random
directed networks," {\em IEEE Trans. Autom. Control}, vol. 51, no. 7,
pp. 1207-1210,  2006.

\bibitem{jad2} A. Tahbaz-Salehi and A. Jadbabaie, ``A necessary and sufficient condition for consensus over
random networks," {\em IEEE Trans. on Autom. Control}, vol. 53, no. 3, pp. 791-795, 2008.

\bibitem{fagnani1} F. Fagnani and S. Zampieri, ``Randomized consensus algorithms over large scale networks," {\it IEEE J. on Selected Areas of Communications}, vol. 26, no.4, pp. 634-649, 2008.

\bibitem{fagnani2}F. Fagnani and S. Zampieri, ``Average consensus with packet drop communication," {\em SIAM J. Control Optim.}, vol. 48, no. 1, pp. 102-133, 2009.


\bibitem{bamieh}S. Patterson, B. Bamieh and A. El Abbadi, ``Convergence rates of distributed average
consensus with stochastic link failures," {\it IEEE Trans.
Autom. Control}, vol. 55, no. 4, pp. 880-892, 2010.

\bibitem{boyd1}S. Boyd,
P. Diaconis and
L. Xiao, ``Fastest mixing markov chain on a graph," {\em SIAM Review}, Vol. 46, No. 4, pp. 667-689, 2004.



\bibitem{gossip2}R. Karp, C. Schindelhauer, S. Shenker, and B. V�cking, ``Randomized
rumor spreading," in {\em Proc. Symp. Foundations of Computer Science},
pp. 564-574, 2000.

\bibitem{gossip1} D. Kempe, A. Dobra, and J. Gehrke, ``Gossip-based computation of aggregate
information," in {\em Proc. Symp. Foundations of Computer Science},
 pp. 482-491, 2003.

\bibitem{boyd} S. Boyd, A. Ghosh, B. Prabhakar and D. Shah, ``Randomized gossip algorithms," {\it IEEE Trans.
Information Theory}, vol. 52, no. 6, pp. 2508-2530, 2006.

\bibitem{f-z}F. Fagnani and S. Zampieri, ``Asymmetric randomized gossip algorithms for consensus," {\em IFAC World Congress}, Seoul, pp. 9051-9056, 2008.



\bibitem{bdo} J. Liu, S. Mou,
A. S. Morse, B. D. O. Anderson,  and
C. Yu, ``Deterministic gossiping," {\em Proceedings of IEEE}, vol. 99, no. 9, pp. 1505-1524, 2011.

\bibitem{murray}J. Lavaei and R.M. Murray,  ``Quantized consensus by means of gossip algorithm," {\em
IEEE Trans. Autom. Control}, vol. 57, no.1, pp. 19-32, 2012.

\bibitem{daron} D. Acemoglu, A. Ozdaglar and A. ParandehGheibi, ``Spread of (Mis)information in social networks," {\em Games and Economic Behavior}, vol. 70, no. 2, pp. 194-227, 2010.

\bibitem{shah}D. Shah, ``Gossip Algorithms," {\em Foundations and Trends in Networking}, Vol. 3, No. 1, pp. 1-125, 2008.


  \bibitem{como} D. Acemoglu, G. Como, F. Fagnani, A. Ozdaglar, ``Opinion fluctuations and persistent disagreement in social networks," in {\em IEEE Conference on Decision and Control}, pp. 2347-2352, Orlando, 2011.


\bibitem{attack1} S. Amin, A. C\'{a}rdenas, and S. Sastry, ``Safe and secure networked
control systems under denial-of-service attacks," in {\em Hybrid Systems:
Computation and Control}, vol. 5469,  pp. 31-45,2009.

\bibitem{attack2} A. A. C\'{a}rdenas, S. Amin, B. Sinopoli, A. Giani, A. A. Perrig, and S. S.
Sastry, ``Challenges for securing cyber physical systems," in {\em Workshop
on Future Directions in Cyber-physical Systems Security}, Newark, NJ,
 Jul. 2009.

\bibitem{attack3} A. Teixeira, S. Amin, H. Sandberg, K. H. Johansson, and S. Sastry,
``Cyber security analysis of state estimators in electric power systems,"
in {\em IEEE Conf. on Decision and Control}, Atlanta,
pp. 5991-5998, 2010.



\bibitem{attack4} F. Pasqualetti and F. Dorfler and F. Bullo, ``Cyber-physical attacks in power networks: models, fundamental limitations and monitor design," In {\em IEEE Conf. on Decision and Control and European Control Conference}, Orlando,  pp. 2195-2201,  2011.

    \bibitem{ft1} A. Tanenbaum, M. van Steen. {\em Distributed
Systems: Principles and Paradigms}. Prentice
Hall, 2002.

\bibitem{ft2} Y. Gershteyn, ``Fault tolerance in distributed systems," {\em IEEE Concurrency}, vol. 4, no. 2, pp. 83-88,  1996.






\bibitem{psl:80} M. Pease, R. Shostak and L. Lamport,   ``Reaching agreement in the presence of faults'',  {\it Journal of the ACM}, vol. 27, no. 2, pp. 228-234, April 1980.

\bibitem{dlpsw:86} D. Dolev, N. A. Lynch, S. S. Pinter, E. W. Stark and W. E. Weihl,  ``Reaching approximate agreement in the presence of faults'',  {\it Journal of the ACM}, vol. 33, no. 3, pp. 499-516, July 1986.







\end{thebibliography}
\end{document}